\documentstyle[12pt]{article}
\begin{document}
\renewcommand{\theequation}{\arabic{section}.\arabic{equation}}

\noindent {\large\bf  EXPANSION IN THE DISTANCE PARAMETER FOR TWO
VORTICES CLOSE TOGETHER } \vspace{1cm}

\hspace{1cm} J. Burzlaff

\vspace{.3cm}

\hspace{1cm} {\it School of Mathematical Sciences}

\hspace{1cm} {\it Dublin City University, Dublin 9, Ireland, and}

\hspace{1cm} {\it School of Theoretical Physics}

\hspace{1cm} {\it Dublin Institute for Advanced Studies, Dublin 4,
Ireland.}

\vspace{.8cm}

\hspace{1cm} E. Kellegher

\vspace{.3cm}

\hspace{1cm} {\it School of Mathematical Sciences}

\hspace{1cm} {\it Dublin City University, Dublin 9, Ireland}

\vspace{2cm}

\noindent {\large \bf Abstract} \vspace{.5cm}

\noindent Static vortices close together are studied for two
different models in 2-dimen- sional Euclidean space. In a simple
model for one complex field an expansion in the parameters
describing the relative position of two vortices can be given in
terms of trigonometric and exponential functions. The results are
then compared to those of the Ginzburg-Landau theory of a
superconductor in a magnetic field at the point between type-I and
type-II superconductivity. For the angular dependence a similar
pattern emerges in both models. The differences for the radial
functions are studied up to third order.

\vspace{2cm}

\noindent DIAS-STP-0014

\newpage

\noindent \section{\large {\bf Introduction}}\vspace{.5cm}

Ever since 't Hooft \cite{one:1} and Polyakov \cite{two:2} found a
monopole solution in the SU(2) Yang-Mills-Higgs theory, solitons
in field theories have been studied extensively. Our understanding
of monopole solutions has been greatly enhanced by an existence
proof for static solutions by Taubes \cite{thr:3} and the
construction of monopole solutions started by Ward \cite{fou:4}.
This process was not matched by quite the same progress in our
understanding of the Abrikosov solutions of the Ginzburg-Landau
theory, although one might have expected that the Abelian Higgs
theory in 2+1 dimensions is actually simpler than the SU(2)
Yang-Mills-Higgs theory in 3+1 dimensions. Again an existence
proof was given by Taubes \cite{fiv:5}. However, only superimposed
vortices can be described explicitly and no explicit construction
of separated vortices is known. In this paper, we want to give the
solution for two vortices close together in terms of an expansion
in the parameters which describe the relative location.

In Sections 2 and 3, we study a model for one complex field. Here
the calculations are simpler than in the Ginzburg-Landau theory
which is our second model. The first model has, however, some
peculiar (unphysical) features. Assuming the most symmetric form
in terms of angular dependence, only two smooth vortices can be
superimposed, and when 'pulled apart', they develop a singularity
at third order. In the Ginzburg-Landau model this does not happen.
In fact, delicate cancellations take place to make the expansion
smooth, at least up to third order. In this model the radial
functions are given as solutions of certain linear ordinary
differential equations. This is discussed in Section 4.

\vspace{.5cm} \noindent \section{\large {\bf Vortex solutions and
zero modes in a simple model}} \vspace{.5cm}

\noindent Our first model is a model \cite{six:6}\cite{sev:7} for
a pair of real fields $\phi^a (\vec{x})$, a,b = 1,2, or
equivalently, for a complex field $\phi = \phi_1+ i \phi_2$. The
Lagrangian density of the model reads
\begin{equation}
{\cal L} = \partial_{[i}{\phi^a} \partial_{j]}{\phi^b} \partial^{[i}{\phi_a}
\partial^{j]}{\phi_b} + (1 - | \phi |^2 )^2 | \phi |^2,
\end{equation}
where $a,b = 1,2$ labels the components of the Higgs field and
$i,j = 1,2$ are the space indices. The square brackets mean
antisymmetrization,
\begin{equation}
{\partial_{[i}\phi^a\partial_{j]}\phi^b} = (\partial_{i}{\phi^a})
(\partial_{j}{\phi^b}) - (\partial_{j}{\phi^a})
(\partial_{i}{\phi^b}) .
\end{equation}
We are working in 2-dimensional Euclidean space, i.e., the space
indices can be raised and lowered without any change in the
formulas. The indices which label the components of the Higgs
field can also be raised and lowered without any change. In terms
of the complex field $\phi$ the Euler-Lagrange equation reads
\begin{equation}
\partial_i\phi^* \partial_j(\partial^{[i} \phi \partial^{j]}\phi^*)
= (1 - | \phi |^2 ) | \phi | \frac{\partial }{\partial\phi} (1 - |
\phi |^2 ) | \phi |.
\end{equation}

Any solution of the equation
\begin{equation}
2\; det \lgroup \frac{\partial\phi^a}{\partial x^{i} } \rgroup =
\pm (1 - | \phi |^2 ) | \phi |
\end{equation} \\
solves the equation of motion (2.3). Note that Eq. (2.4) is a
first order equation whereas Eq. (2.3) is of second order. So we
would expect that (2.4) is somewhat easier to solve than (2.3).
For different types of models, this reduction of order was first
introduced by Bogomolnyi \cite{eig:8}. That is why we call Eq.
(2.4) the Bogomolnyi equation here. Any solution of (2.4) also
attains the lower bound in the following inequality,
\begin{equation}
A = \int_{{\bf R}^2} {\cal L}\;d^2 x \geq \frac{16 \pi}{15} |Q|,
\;\;\;\;\;\;\;\; \;\;
\end{equation}
where
\begin{equation}
Q = \frac{15}{8 \pi} \int_{{\bf R}^2} i  \epsilon_{ij} (1 - | \phi
|^2 ) | \phi | (\partial^i \phi ) (  \partial^j \phi^{*}) d^2 x
\end{equation}
is the winding number. Finally, all finite-action solutions
actually solve the Bogomolnyi equation, so we do not miss out on
any by concentrating on the first order equation.

We now seek to attain a smooth finite-action solution of Eq.
(2.4). For
\begin{equation}
\phi = f(r) e^{i n \theta}
\end{equation}
Eq. (2.4) reduces to
\begin{equation}
\frac{n f(r) f^{'}(r)}{r} = \frac{1}{2}( 1 - f^2 ) f.
\end{equation}
Since $f \rightarrow 0$ as $r \rightarrow 0$ (otherwise $\phi$ in
(2.7) is not defined at the origin), we have
\begin{equation}
f = \tanh \frac{r^2}{4 n}.
\end{equation}

The solution $\phi$ in (2.7) with f(r) given by (2.9) is defined
in the whole of ${\bf R}^2$ and is clearly a $C^\infty$ function
in ${\bf R}^2 \setminus \{0\}$. Since
\begin{eqnarray}
f \approx 1 - 2 \exp \frac{ r^{2} }{ 2 n }  \;\;\;\; {\rm as}
\;\;\;\; r \rightarrow \infty ,
\end{eqnarray}
$\phi$ has the right asymptotic behaviour for a solution with
winding number n. We still have to ensure that $\phi$ is
$C^\infty$ at the origin. There we use the Taylor expansion of
$f$,
\begin{eqnarray}
f = \sum_{K=1}^{\infty} \frac{2^{2 k} ( 2^{2 k} - 1) B_{2 k} }{(2k) !}
( \frac{r^2}{4 n} )^{2k - 1}
= \frac{r^2}{4 n} - \frac{1}{3}(\frac{r^2}{4 n})^3 + ...
\end{eqnarray}
where $B_k$ is the $k^{th}$ Bernoulli number. We see that for $n =
2$ and only for $n = 2$, $\phi$ is a polynomial in $x^i$. In this
model, we have the (somewhat peculiar) situation that within the
most natural ansatz (2.7) smooth finite action solutions exist
only for $n = 2$, i.e., we only have a solution of the form (2.7)
for 2 vortices.

We have found the solution for two vortices sitting on top of each
other, which we now denote by $\hat{\phi}$. To extend our study to
two vortices slightly apart we consider $\phi = \hat{\phi} +
\gamma$, where $\gamma$ is very small, and we solve the Bogomolyni
equation, linearized in $\gamma$. Equation (2.4) becomes
\[
(f^{'}\cos \theta \cos 2 \theta   + \frac{2}{r} f \sin \theta \sin
2 \theta) \frac{\partial \gamma^2}{\partial x^2} +(f^{'}\sin
\theta \sin 2 \theta   + \frac{2}{r} f \cos \theta \cos 2 \theta)
\frac{\partial \gamma^1}{\partial x^1}
\]
\[
- (f^{'}\sin \theta \cos 2 \theta   - \frac{2}{r} f \cos \theta
\sin 2 \theta) \frac{\partial \gamma^2}{\partial x^1} - (f^{'}\cos
\theta \sin 2 \theta   - \frac{2}{r} f \sin \theta \cos 2 \theta)
\frac{\partial \gamma^1}{\partial x^2} \nonumber
\]
\begin{equation}
 = \frac{1}{2} (1 - 3 f^2) \;\;(\gamma^1 \cos 2 \theta + \gamma^2
\sin 2 \theta )
\end{equation}
We find a 2-parameter family of zero modes,
\begin{eqnarray}
\gamma(r) = [\alpha + \beta + \imath(\alpha - \beta) ] h(r)
\;\;\;\; {\rm with} \;\;\;\; h(r) = \frac{ \sinh \frac{r^{2}}{8}
}{ \cosh^{3} \frac{r^{2}}{8} }.
\end{eqnarray}\\
These zero modes are $C^{\infty}$ functions which vanish
exponentially at infinity. By a rotation, one of the parameters
could be removed and the vortices could be positioned on the
$x$-axis, say. Since this does not simplify the calculations
significantly, we will retain both parameters. Retaining the two
parameters would also be necessary for a study of vortex
scattering in the slow-motion approximation. This study is not
done in this paper.

\vspace{.5cm} \noindent \section{\large {\bf The quadratic and
cubic terms}} \setcounter{section}{3} \setcounter{subsection}{3}
\setcounter{equation}{0} \vspace{.5cm}

\noindent We now consider $\phi = \hat{\phi} + \gamma + \delta$,
and equate the second order terms in the Bogomolyni equation
(2.4). This leads to the equation
\begin{eqnarray}
\frac{2}{r} ( f^{'} \cos 2 \theta \frac{\partial \delta^2}{\partial \theta}
+ 2 f \sin 2 \theta \frac{\partial \delta^2}{\partial r}
- f^{'} \sin 2 \theta \frac{\partial \delta^1}{\partial \theta}
+ 2 f \cos 2 \theta \frac{\partial \delta^1}{\partial r} )
\nonumber \\
= (\alpha^2 + \beta^2) f h^2 ( \frac{1}{f^2} - 3)
- \frac{1}{2} f h^2 ( 3 + \frac{1}{f^2}
) \; [\alpha^2 (\cos 2 \theta  + \sin 2 \theta)^2
\nonumber \\
+ 2 \alpha \beta (\cos^2 2 \theta  - \sin^2 2 \theta )
+ \beta^2 (\cos 2 \theta  - \sin 2 \theta) ^2 ]
\nonumber \\
+ (1 - 3 f^2) (\delta^1 \cos 2 \theta + \delta^2 \sin 2 \theta)
\end{eqnarray}
with $f(r)$ given in (2.9) and $h(r)$ given in (2.13).

With $\delta$ of the form
\begin{equation}
\delta = \alpha^2 F(r, \theta) + 2 \alpha \beta G(r, \theta) +
\beta^2 H (r, \theta),
\end{equation}
we obtain the following equation for $F(r, \theta)$,
\begin{eqnarray}
\frac{2}{r} ( f^{'} \cos 2 \theta \frac{\partial F^2}{\partial
\theta} + 2 f \sin 2 \theta \frac{\partial F^2}{\partial r} -
f^{'} \sin 2 \theta \frac{\partial F^1}{\partial \theta} + 2 f
\cos 2 \theta \frac{\partial F^1}{\partial r} ) \nonumber \\ = h^2
(\frac{1}{f} - 3 f) - \frac{h^2}{2}  (3 f + \frac{1}{f}) (\cos 2
\theta + \sin 2 \theta)^2 \nonumber \\ + (1 - 3 f^2) (F^1 \cos 2
\theta + F^2 \sin 2 \theta) .
\end{eqnarray}
To solve this equation we seek a solution of the form
\begin{equation}
F = f_1 (r) \exp^{\imath 2 \theta} - \imath f_2 (r) \exp^{- \imath
2 \theta} .
\end{equation}

The ansatz (3.4) leads to two decoupled equations for $f_1$ and
$f_2$. In terms of the variable $\xi = r^2 /8$, they read
\begin{equation}
\frac{d f_1}{d \xi} + \frac{1}{f} (3 f^2 -1 - \frac{d f}{d \xi}) f_1
=
\frac{h^2}{2 f^2}(1 - 9 f^2),
\end{equation}
\begin{equation}
\frac{d f_2}{d \xi} + \frac{1}{f} (3 f^2 -1 + \frac{d f}{d \xi} ) f_2
=
-\frac{h^2}{2 f}(1 + 3 f^2).
\end{equation}

The general solutions to equation (3.5) is
\begin{equation}
f_1 = \frac{1}{\cosh^2 \xi} ( \frac{3 \sinh \xi}{2 \cosh^3 \xi} -
\frac{\sinh \xi}{\cosh \xi} + C_1 )
\end{equation}
The function $f_1$ is a $C^\infty$ function for $0 < \xi <
\infty$. For $\xi \rightarrow 0$, $f_1 \rightarrow C_1$ holds.
This implies that $C_1 = 0$; otherwise $F$ in (3.4) is not defined
at the origin. Therefore, $f_1$ reads
\begin{equation}
f_1 = \frac{3 \sinh \xi}{2 \cosh^5 \xi} - \frac{\sinh \xi}{\cosh^3
\xi} .
\end{equation}
\\
The expansion of $f_1$ near the origin is of the form
\begin{equation}
f_1 = \sum_{k = 1}^{\infty} a_k \xi^k = \sum_{k = 1}^{\infty} a_k
(\frac{r^2}{8} )^k .
\end{equation}
\\
Hence, the first term in (3.4) is a $C^\infty$ function of $x^1$
and $x^2$ at the origin. We also see that $f_1$ vanishes
exponentially at infinity. So its contribution to $\phi$ does not
change the winding number (2.6) which is a multiple of the action.

A similar calculation yields a one parameter family of solutions
to Eq. (3.6), namely
\begin{equation}
f_2 = \frac{\sinh \xi}{2 \cosh^3 \xi} - \frac{3 \sinh^3 \xi}{2
\cosh^5 \xi} + C_2 \frac{\sinh ^2 \xi}{\cosh ^4 \xi}.
\end{equation}
In contrast to $f_1$, all the solutions $f_2$ are acceptable. In
fact, for all $C_2$, $f_2$ is of the form
\begin{equation}
f_2 = \sum_{k=1}^\infty b_k \xi^k = \sum_{k=1}^\infty b_k (\frac{r^2}{8} )^k
\end{equation}
near the origin, and therefore the second term in (3.4) is in
$C^\infty( {\bf R}^2 )$. The winding number and the action are
also not altered because $f_2$ decays exponentially at infinity.

The functions $G$ and $H$ in (3.2) can be found in the same way.
If we put all results together, we obtain the second order terms,
\begin{equation}
\delta = (\alpha^2 + \beta^2) f_1 (r) \exp^{\imath 2 \theta} +
\imath(\alpha - i \beta)^2 f_2 (r) \exp^{- \imath 2 \theta} ,
\end{equation}
where $f_1$ and $f_2$ are given by (3.8) and (3.10), respectively.

To find the cubic terms, we consider $\phi = \hat{\phi} + \gamma +
\delta + \epsilon$, with $\gamma$ given in (2.13) and $\delta$
given by (3.12). We set $\beta =0$ and concentrate on
\begin{equation}
\epsilon = \alpha ^3 I(r,\theta ) .
\end{equation}
For the Bogomolnyi equation to hold, $I$ must satisfy
\[
\frac{2}{r} ( f^{'} \cos 2 \theta \frac{\partial I^2}{\partial
\theta} + 2 f \sin 2 \theta \frac{\partial I^2}{\partial r} -
f^{'} \sin 2 \theta \frac{\partial I^1}{\partial \theta} + 2 f
\cos 2 \theta \frac{\partial I^1}{\partial r} ) \nonumber
\]
\[
+ h^{'} ( 2 f_1 \cos 2 \theta + 2 f_2 \sin 2 \theta) + h^{'} ( 2
f_1 \sin 2 \theta + 2 f_2 \cos 2 \theta)
\]
\[
= - 3 f^2 (I^1 \cos 2 \theta  + I^2 \sin 2 \theta) - f_2 (\cos 2
\theta + \sin 2 \theta )
\]\[
- 3 f h (\cos 2 \theta  + \sin 2 \theta ) (f_1 - 2 f_2 \cos 2
\theta \sin 2 \theta ) - 3 (\cos 2 \theta  + \sin 2 \theta ) h^2
\]
\[
+(I^1 \cos 2 \theta  + I^2 \sin 2 \theta) + \frac{h}{2}[f_1 (\cos
2 \theta + \sin 2 \theta ) -f_2(\cos 2 \theta  + \sin 2 \theta ) ]
\]
\[
+ \frac{h^3}{2}(\cos 2 \theta + \sin 2 \theta )^3
-\frac{h^3}{f^2}(\cos 2 \theta  + \sin 2 \theta )
\]
\begin{equation}
- \frac{1}{f} (\cos 2 \theta  + \sin 2 \theta ) (f_1-2f_2\cos 2
\theta  \sin 2 \theta ) + \frac{1}{2 f^2} (I^1 \cos 2 \theta + I^2
\sin 2 \theta)^3
\end{equation}

To solve equation (3.14) we seek a solution of the form
\[
I^1 = g_1 (\xi) + g_2(\xi) (\cos 4 \theta -\sin 4 \theta) ,
\]
\begin{equation}
I^2 = g_1 (\xi) - g_2(\xi) ( \cos 4 \theta + \sin 4 \theta) .
\end{equation}
This implies that $g_1$ and $g_2$ must satisfy the equations
\begin{equation}
\frac{d g_1}{d \xi} + (3f - \frac{1}{f}) g_1
=
-  \frac{f_1 + f_2}{f}\frac{dh}{d\xi}  - 6 h f_1 + \frac{9}{2} h
f_2 - \frac{h f_2}{2 f^2} - \frac{h^3}{4 f^3} - \frac{9 h^3}{4 f},
\end{equation}
\begin{equation}
\frac{d g_2}{d \xi} - ( \frac{1}{f} - 3 f + \frac{2}{f}
\frac{df}{d\xi} ) g_2
=
-\frac{h f_2}{2 f^2} - \frac{3 h f_2}{2} - \frac{h^3}{4 f^3} -
\frac{h^3}{4 f}.
\end{equation}

The general solution to Eq. (3.17) is
\begin{equation}
g_2 = \frac{\sinh \xi }{4 \cosh^5 \xi} - \frac{5 \sinh^3 \xi}{4
\cosh^7 \xi} + C_2 \left( \frac{\sinh ^2 \xi }{2 \cosh^4 \xi} -
\frac{3 \sinh^4 \xi}{2 \cosh^6 \xi} \right) + C_3 \frac{\sinh ^3
\xi }{\cosh^5 \xi} .
\end{equation}
All solutions (3.18) decay exponentially at infinity. For
$r\rightarrow 0$, however,
\begin{equation}
g_2 (r) = \frac{1}{24} r^2 + \ldots
\end{equation}
Hence, $I$ in (3.15) is not a $C^\infty$ function on ${\bf R}^2$.
Our expansion gets singular at third order for the ansatz (3.15).
In the next section we will discuss a realistic model in which a
similar pattern emerges but no singularities occur.

\vspace{.5cm} \noindent \section{\large {\bf Abrikosov vortices}}
\setcounter{section}{4} \setcounter{subsection}{4}
\setcounter{equation}{0} \vspace{.5cm}

The Ginzburg-Landau theory of a superconductor in a magnetic field
in direction \(z\) is given by the Lagrangian density
\begin{equation}
{\cal L}= \frac{1}{4}F_{ij} F^{ij} + \frac{1}{2}(D_i \phi)(D^i
\phi)^* + \frac{\lambda}{8}(\mid \phi \mid \; ^{2} -1)^{2},
\end{equation}
where \(\phi\) is the complex Higgs field, and \(D_i \phi =
\partial_i \phi - i A_i \phi\) and \(F_{ij} = \partial_i A_j
\partial_j A_i\) in terms of the gauge potentials \(A_i,\; i=1,2\).
The Euler-Lagrange equations are
\begin{equation}
D_iD^i \phi = \frac{\lambda}{2} \phi (1- \mid \phi \mid
\;^2),\;\;\;\;\;\;
\partial_i F^{ij} = \frac{\imath}{2}[\phi (D^j \phi) ^*- \phi^* D^j \phi]
\end{equation}

In the special case \(\lambda=1\) it can be shown \cite{nin:9}
that all finite action solutions of Eq. (4.2) satisfy the
first-order Bogmolnyi equations \cite{eig:8},
\begin{equation}
F_{12} = \frac{1}{2}(1-\mid \phi \mid \;^2), \;\;\;\;\;\; D_1 \phi
= -iD_2 \phi .
\end{equation}
It has also been shown \cite{nin:9} that a \(2n\)-parameter family
of solution of (4.3) exists with winding number
\begin{equation}
n=\frac{1}{2\pi} \int_{{\bf R}^2} F_{12}\; d^{2} x .
\end{equation}
This family describes \(n\) vortices sitting at \(n\) position in
space.

Even for \(n\) vortices sitting on top of each other, the solution
is not known explicitly in terms of elementary functions.  It is
known \cite{ten:10}, however, that this solution is of the form
\begin{equation}
\phi = f (r) e^{\imath n \theta}, \;\;\;\;\;\; A_i = - \frac{n
a(r)}{r^2}\varepsilon_{ij} x^j ,
\end{equation}
where \(f\) and \(a\) satisfy
\begin{equation}
rf^ \prime - n (1-a) f = (2n/r) a ^ \prime + f^2 - 1 = 0
\end{equation}
and
\begin{equation}
f(0)=a(0)=0, \;\;\;\; \lim_{r \rightarrow \infty}f(r)=\lim_{r
\rightarrow \infty} a(r) = 1.
\end{equation}

In the following, we restrict our attention to \(n=2\) and use the
solution (4.5) as the zero order term in an expansion in the
separation parameters.  The first order terms are given by the two
zero modes describing the separation of the votices.  These were
found by Weinberg \cite{ele:11}.  Using his results we can write,
up to quadratic terms,
\begin{equation}
\phi = fe^{2\imath \theta} + 2 (\alpha + \imath \beta)k f +
\alpha^2 \psi + \alpha \beta \phi + \beta^2 \chi + \ldots,
\end{equation}
\[ A_1 + \imath A_2 = \imath \frac{2a}{r} e^{\imath \theta} - 2\imath (\alpha +
\imath \beta) (k^ \prime + \frac{2k}{r})e^{-\imath\theta} \]
\begin{equation}
+ \alpha^2 ( B_1 + \imath B_2) + \alpha\beta (C_1 + \imath C_2) +
\beta^2 (D_1 + \imath D_2) + \ldots
\end{equation}
Here the radial function \(k(r)\) satisfies
\begin{equation}
k^{\prime \prime} + \frac{1}{r}k^ {\prime} - (f^2 + \frac{4}{r^2})
k=0,
\end{equation}
with
\begin{equation}
\lim_{r \rightarrow 0} r^2 k=1, \;\;\;\;\;\; \lim_{r \rightarrow
\infty} k(r) = 0.
\end{equation}
Our task is to determine \(\psi, \phi,\chi,B_i, C_i, D_i\), which
are functions of \(r\) and \(\theta\).

Equating the \(\alpha^2\)-terms in the Bogomolnyi equations (4.3),
we obtain
\begin{equation}
(\partial_1 + \imath\partial_2)\psi+\frac{2a}{r}\psi e^{\imath
\theta} - \imath f(B_1 + \imath B_2 )e^{2\imath \theta} = 4kf(k^
\prime + \frac{2k}{r})e^{-\imath\theta} ,
\end{equation}
\begin{equation}
\partial_1 B_{2}-\partial_{2}B_{1} + \frac{1}{2}(f \psi e^{-2\imath \theta}
+ f \psi e^{2\imath \theta}) = - 2k^2 f^2 .
\end{equation}
A Fourier expansion with the minimal number of nonzero terms leads
to the ansatz
\[\psi= g(r)f(r)e^{2\imath \theta} + \tilde{g}(r)e^{-2\imath \theta},\]
\begin{equation}
B_1 + \imath B_2 = \tilde{b}(r)e^{\imath \theta} + \imath
b(r)f(r)e^{-3\imath \theta},
\end{equation}
and to equations for \(g(r), \tilde{g}(r), b(r)\) and \(
\tilde{b}(r).\)  The equations for \( \tilde{g}(r)\) and
\(\tilde{b}(r)\) read
\begin{equation}
\tilde{g}=\frac{1+2a}{r} b-b^ \prime, \;\;\;\;\;\;
\tilde{b}=-\imath h^ \prime .
\end{equation}
The functions \(g(r)\) and \(b(r)\) must satisfy the equations
\begin{equation}
g^ {\prime \prime} + \frac{1}{r} g^ \prime - f^2 g = 2k^2 f^2,
\end{equation}
\begin{equation}
b^{\prime \prime} + \frac{1}{r} b^ \prime -
(\frac{1+f^2}{2}+\frac{1+4a+4a^2}{r^2}) b=-4kf(k^ \prime +
\frac{2k}{r}).
\end{equation}

Equating the \(\alpha \beta\)-terms and the \(\beta^2\)-terms in
the Bogonolnyi equation (4.3), we obtain equations for \(\phi\)
and \(C_i\), and for \(\chi\) and \(D_i\) respectively.  These
equations, which are very similar to equations (4.12) and (4.13),
can again be solved by functions with the same
\(\theta\)-dependence as in (4.14) but with slightly different
radial functions. Collecting all results, we can write the
solution, up to quadratic terms, in the form
\[
\phi = fe^ {2\imath\theta} + 2( \alpha + \imath \beta)k f
\]
\[
+ (\alpha^2 + \beta^2) g f e ^{2\imath \theta} + (\alpha + \imath
\beta)^2 ( \frac{1+2a}{r} b-b^{\prime})e^{-2\imath \theta}+ \ldots
\]
\[
A_1 + \imath A_2 = \imath \frac{2a}{r}e^{\imath \theta} - 2\imath
(\alpha + \imath \beta)(k^{\prime} + \frac{2k}{r})e^{-\imath
\theta}
\]
\begin{equation}
-\imath(\alpha^2 + \beta^2)g^{\prime}e^{\imath \theta} + \imath
(\alpha + \imath \beta)^2 b f e^{-3\imath \theta}+ \ldots
\end{equation}

It remains to be shown that the quadratic terms in (4.18) are
\(C^{\infty}\) functions on \({\bf R}^2\) which do not change the
action (and the winding number). To this end we use the asymptotic
expansions of \(f, a\) and \(k\) at zero \cite{twe:12},
\begin{equation}
f(r)=f_1 r^2 + \frac{1}{8} f_1 r^4 +  \ldots, \;\; a(r)=
\frac{1}{8} r^2 - \frac{1}{24} f_1^2 r^6+ \ldots, \;\; k(r)=r^{-2}
+ k_1 r^2 + \ldots ,
\end{equation}
where $f_1 = .236$ and $k_1 = -.025$ from the numerical analysis.
We find that the solutions of (4.16) and (4.17) have the following
expansions at the origin,
\[
g(r)= g_{-1} \log r + g_1 + \frac{1}{2} f_1^2 r^2 + \ldots
\]
\begin{equation}
b(r)=b_{-1}r^{-1}+b_1 r + (\frac{1}{8} b_1 - 2 f_1 k_1 ) r^3 +
\ldots
\end{equation}
The higher order terms in \(g(r)\) are even powers of \(r\),
whereas the higher order term in \(b(r)\) are odd powers of \(r\).
Hence, the quadratic terms in (4.18) are \(C^{\infty}\) near the
origin if and only if \(h_{-1} = b_{-1} = 0.\) So far the
constants $g_1$ and $b_1$ are arbitrary.

For large \(r\) the functions \(f, a,\) and \(k\) have the
following asymptotic behavior \cite{twe:12}:
\[
f(r)=1+{\tilde f}_1 (r)e^{-r} + \ldots ,
\]
\begin{equation}
a(r)=1+{\tilde a}_1 (r)e^{-r} + \ldots ,
\end{equation}
\[
k(r)={\tilde k}_1 (r) e^{-r} + \ldots ,
\]
with coefficient functions which are polynomially bounded.  This
leads to the existence of exponentially decaying solutions which
asymptotically are of the form
\begin{equation}
g(r) = {\tilde g}_1 (r) e^{-r} + \ldots, \;\;\;\; b(r)= {\tilde
b}_1 (r) e^{-r} + \ldots
\end{equation}
Here ${\tilde g}_1$ and ${\tilde b}_1$ are polynomially bounded.

By numerical integration, the coefficients $g_1$ and $b_1$ which
lead to an exponential fall-off at infinity, are found to be $g_1
= -.144$ and $b_1 = -.026$. The existence of such functions can be
explained analytically as follows: Equation (4.16) shows that for
positive $g_1$, g cannot have a maximum for any $r$. So the
function diverges exponentially. For very small $g_1$, the term on
the right-hand side of (4.16) will force the function to cross the
$r$-axis, and then, as before, diverge exponentially. For very
large negative $g_1$, the third term in (4.16) will force $g$ to
go through a maximum for large $r$. After that, the function
cannot have a minimum and must go to minus infinity. Because of
the continuous dependence on the initial data, we have an open set
of data for which $g$ crosses the $r$-axis, and an open set of
data for which $g$ goes through a maximum below the $r$-axis.
Therefore, we have at least one value of $g_1$ for which the
function does neither. This function must converge and does so to
zero, exponentially.

A similar argument explains the existence of an acceptable
solution $b(r)$ to Eq. (4.17). The right-hand side of that
equation is positive. So again $b$ cannot have a maximum above the
$r$-axis. Also, for very small negative $b_1$, the right-hand side
will force $b$ to go through a minimum and then cross the
$r$-axis. For very large negative $b_1$, the third term in (4.17)
prevents $b$ from going through a minimum. In between these two
possibilities we find the desired solution which goes through a
minimum but does not cross the $r$-axis. Such a solution must
decay exponentially.

The cubic terms can be calculated in the same manner. We find, at
third order,
\[
\phi = \ldots + (\alpha + \imath\beta)(\alpha^2 + \beta^2)f h +
(\alpha + \imath\beta)^3 (-c' + \frac{3+2a}{r} c)
e^{-4\imath\theta} + \ldots ,
\]
\[
A_1 + \imath A_2 = \ldots
\]
\begin{equation}
+ \imath (\alpha + \imath\beta)(\alpha^2 + \beta^2) [-h'-
\frac{2}{r} h + 2g(k' +\frac{2k}{r}) + 2kg'] e^{-\imath\theta} +
\imath (\alpha + \imath\beta)^3 f c e^{-5\imath\theta} + \ldots
\end{equation}
The new radial functions, $h(r)$ and $c(r)$, satisfy the
equations,
\begin{equation}
h'' + \frac{1}{r} h' - (f^2 + \frac{4}{r^2} ) h = 4k'g' + 2fk
(2fk^2 +3fg + \frac{1+2a}{r} b - b'),
\end{equation}
\begin{equation}
c'' + \frac{1}{r} c' - (\frac{1+f^2}{2} + \frac{9+12a+4a^2}{r^2} )
c = 2kf^2 b -2(k' +\frac{2k}{r} ) (\frac{1+2a}{r} b - b').
\end{equation}

Near the origin, Eq. (4.25) has a series solution in powers of
$r^2$ of the form
\begin{equation}
h(r) = f_1^2 + h_1 r^2 + h_2 r^4 + \ldots
\end{equation}
The constant term is given in terms of the coefficient $f_1$ of
the leading term in the expansion (4.19) of $f(r)$. The form of
this term leads to the cancellation of the $r^{-1}$-terms in the
radial function multiplying $e^{-\imath \theta}$ in (4.24), and
thus ensures that this term in (4.23) is $C^{\infty}$ on ${\bf
R}^2$. The series in odd powers of $r$ for $c(r)$ which solves Eq.
(4.25) near the origin, is
\begin{equation}
c(r) = c_1 r^3 + c_2 r^5 + \ldots
\end{equation}
The form of the series solutions at the origin guarantees that the
cubic terms in (4.23) are $C^\infty$ functions on ${\bf R}^2$. For
large $r$, Eqs (4.24) and (4.25) have exponentially decaying
solutions.

\vspace{.5cm} \noindent \section{\large {\bf Conclusions}}
\vspace{.5cm}

\noindent Our expansions show a simple $\theta$-dependence in
terms of trigonometric functions. In both models, the expansion of
$\phi$ exhibits the following pattern:
\[
\matrix{&&&&&&e^{2\imath\theta}\cr
         &&&&&e^{0\imath\theta}&\cr
         &&&&e^{-2\imath\theta}&&e^{2\imath\theta}\cr
         &&&e^{-4\imath\theta}&&e^{0\imath\theta}&\cr
         &&e^{-6\imath\theta}&&e^{-2\imath\theta}&&e^{2\imath\theta}\cr
         &e^{-8\imath\theta}&&e^{-4\imath\theta}&&e^{0\imath\theta}\cr
         \ldots&&\ldots&&\ldots&&\ldots}
\]
Here the first line gives the $\theta$ dependence of the zero
order term; the second line gives the first order term, and so on.
We get a similar triangular pattern for the $\theta$ dependence of
$A_1 +\imath A_2$ at any order. For the radial functions we find
differences between the two models. In the model for one complex
field, the radial functions can be given explicitly in terms of
exponential functions. However, for the angular dependence (3.15),
a singularity occurs at the origin. (We have found no solution to
(3.14) which is not of the form (3.15); we have no proof that
there is none.)

For the Ginzburg-Landau theory on the other hand, the expansion is
smooth, at least up to the order to which we carried out our
calculations. In this model the radial functions are not given in
terms of well-known functions. Having used the technique to
calculate the terms up to third order, it is quite clear how to
proceed to any order, and also how to proceed in the case of more
than two vortices. We expect these expansions to converge for
small separation parameters in the physical Ginzburg-Landau model.
However, we do not have an estimate of the radius of convergence.

\end{document}